\documentclass[manuscript]{aastex}

\usepackage[latin1]{inputenc}
\usepackage[T1]{fontenc}
\usepackage[english]{babel}
\usepackage{amsfonts}
\usepackage{eucal}
\usepackage{amsthm}
\usepackage{amsbsy}
\usepackage{amssymb}
\usepackage{amsmath}
\usepackage{wasysym}
\usepackage{graphicx}
\usepackage{babel}
\usepackage{subfigure}
\usepackage{natbib}
\usepackage{fancyhdr}
\usepackage{float}
\usepackage{multicol}
\usepackage{textcomp}
\usepackage{epsfig}
\usepackage{ mathrsfs }
\begin{document}


\title{0.94 - 2.42  $\mu$m ground-based transmission spectra of the hot Jupiter HD-189733b}


\author{C. Danielski}
\affil{Dept. of Physics \& Astronomy, University College London, Gower Street, WC1E 6BT, UK}
\email{camilla@star.ucl.ac.uk}

\author{P. Deroo}
\affil{Jet Propulsion Laboratory, California Institute of Technology, 4800 Oak Grove Drive, Pasadena, California 91109-8099, USA}

\author{I. P. Waldmann, M. D. J. Hollis, G. Tinetti}
\affil{Dept. of Physics \& Astronomy, University College London, Gower Street, WC1E 6BT, UK}

\author{M. R. Swain}
\affil{Jet Propulsion Laboratory, California Institute of Technology, 4800 Oak Grove Drive, Pasadena, California 91109-8099, USA}


\begin{abstract}
We present here new  transmission spectra of the hot Jupiter HD-189733b using the SpeX instrument 
on the NASA Infrared Telescope Facility. We obtained two nights of observations where we recorded the primary transit of the planet
in the J-, H- and K-bands simultaneously, covering a spectral range from 0.94 to 2.42 $\mu$m.
We used Fourier analysis and  other de-trending techniques validated previously  on other datasets to clean the data.
We tested the statistical significance of our results by calculating the auto-correlation function, and we found that, after the detrending, 
auto-correlative noise is diminished at most frequencies. 
Additionally, we repeated our analysis on the out-of-transit data only, showing that the residual telluric contamination is well within the error bars.
While these techniques are very efficient when multiple nights of observations are combined together, our results prove that even one good night of observations is enough to provide 
statistically meaningful data. 
Our observed spectra are consistent with space-based data recorded in the same wavelength interval by multiple instruments, 
indicating that ground-based facilities are becoming a viable 
and complementary option to spaceborne observatories. The best fit to the features in our data was obtained with water vapor. Our error bars are 
not small enough to address the presence of additional molecules, however by combining the information contained in other datasets with our results, 
it is possible to explain all the available observations 
with a modelled atmospheric spectrum containing water vapor, methane, carbon monoxide and hazes/clouds.

\end{abstract}




\section{Introduction}

With over 900 exoplanets discovered and more to come in the near future, 
exoplanet research today is anything but dull. 
The synergy between space and ground work is delivering outstanding results and the menagerie of exoplanets is becoming more diverse.
For instance the \textit{Kepler} mission (Borucki et al. 2010, 2011) has delivered, among other impressive discoveries,
the smallest exoplanet known to date (Barclay et al. 2013), and the high-resolution spectrograph \textit{HARPS} at the
European Southern Observatory has
detected a new Earth-sized exoplanet around $\alpha$ Centauri B (Dumusque et al. 2012).

Beyond the key parameters of mass, radius and orbital inclination, the next critical step is to determine the chemical composition of these exotic worlds.
Detecting atmospheric features, which have a contrast of about 10$^{-4}$ compared to the host 
star's radiation, is quite a challenge. However, for transiting exoplanets this has proven to be feasible from space and the ground.
With the high stability of the Spitzer Space Telescope and the Hubble Space Telescope, the spectra of bright close-in massive planets have been obtained
and ionic/atomic/molecular species such as ionized forms of hydrogen, silicon, carbon, magnesium, as well as alkali metals, water vapor, methane, carbon monoxide and dioxide 
have been detected in a handful of these planets.
With the demise of \ \textquotedblleft cold Spitzer\textquotedblright \ and NICMOS though, the wavelength range covered from space has narrowed. 
However on the positive side the new Wide Field Camera 3 
on Hubble is already delivering novel  and interesting results (Berta et al. 2012, Swain et al. 2013).

Whilst in previous years the transit technique, combined with 
the photometric precision of Hubble and Spitzer, has been an asset for the success of this field, 
the impossibility of repeating observations had led to debates in the community.
Some of these concerns have been addressed by adopting more robust and objective statistical techniques to remove instrumental 
systematics (Gregory 2011, Gibson et al. 2012a, Waldmann 2012, Waldmann et al. 2012).
Furthermore, there has been a rapid escalation of successful results from the ground over the last few years (Redfield et al. 2008, Snellen et al. 2008, 2010, 
Bean et al. 2010, Swain et al. 2010, Waldmann et al. 2012, Brogi et al. 2012).
Ground-based observations have the non-trivial limitation of having telluric line contamination interfering with 
measurements, especially in the infrared where most of the key molecules show stronger absorption features.
However, observations from the ground can be repeated more easily (Waldmann et al. 2012) and in some cases they can cover
spectral regions not reachable from space (Swain et al. 2010), or provide higher spectral resolution (Snellen et al. 2010, Mandell et al. 2011, Brogi et al. 2012, De Kok et al. 2013, Birkby et al. 2013).\\
The hot Jupiter HD-189733b has been the most observed planet to date due to the brightness of its mother star
and its favourable atmosphere. Observations, taken with multiple instruments and by different teams (e.g. Knutson et al. 2007, Tinetti et al. 2007, Charbonneau et al. 2008, 
Grillmair et al. 2008, Swain et al. 2008, 2009), have been interpreted with 
the presence of H$_{2}$O, CH$_{4}$, CO$_2$ and CO in its atmosphere (Barman 2008, Swain et al. 2009, Madhusudhan $\&$ Seager, 2009, Tinetti et al. 2010a, 
Lee et al. 2012, Line et al. 2012, De Kok et al. 2013, Birkby et al. 2013).
Gibson et al. 2012b and Pont et al. 2013 suggest instead that hazes extend from the UV down to the MIR at the terminator.
The recent detection of strong feature at 3.3 $\mu$m in its dayside (Swain et al. 2010, Waldmann et al. 2012) has been explained as non-thermal methane emission from the planet. This interpretation has been debated in the literature (Mandell et al. 2011).

Here we present new observations of HD-189733b in the J-, H- and K-bands with the NASA Infrared Telescope
Facility (IRTF)/SpeX instrument. 
While only two nights out of three could be used for our analysis, the remaining nights were sufficiently photometrically stable to allow the extraction of low resolution spectra in those bands. 
Given that the J-, H- and K-bands were observed simultaneously, these measurements have a clear advantage for atmospheric interpretation over photometric data recorded at different times
(Sing et al. 2009, Gibson et al. 2012b). The activity of the star (Knutson et al. 2007, 2009, Agol et al. 2010) in fact may prevent the level of accuracy needed to detect molecular 
features in the planetary atmosphere (Ballerini et al. 2012).

To analyse the data,  we have used the techniques described in Swain et al. (2010) and Waldmann et al. (2012) tailored for transmission spectroscopy. 
These techniques have been validated on observations with a reference star in the field over the 0.28 - 4.2 $\mu$m spectral range (L-band), where the telluric contamination is high.
Waldmann et al. (2012) found that the spectrum of the observed reference star was flat, indicating an adequate background subtraction.

\section{Data reduction and spectral extraction}
 \subsection{\textit{Observations}}

\begin{figure}[f]
\epsscale{0.55}
\plotone{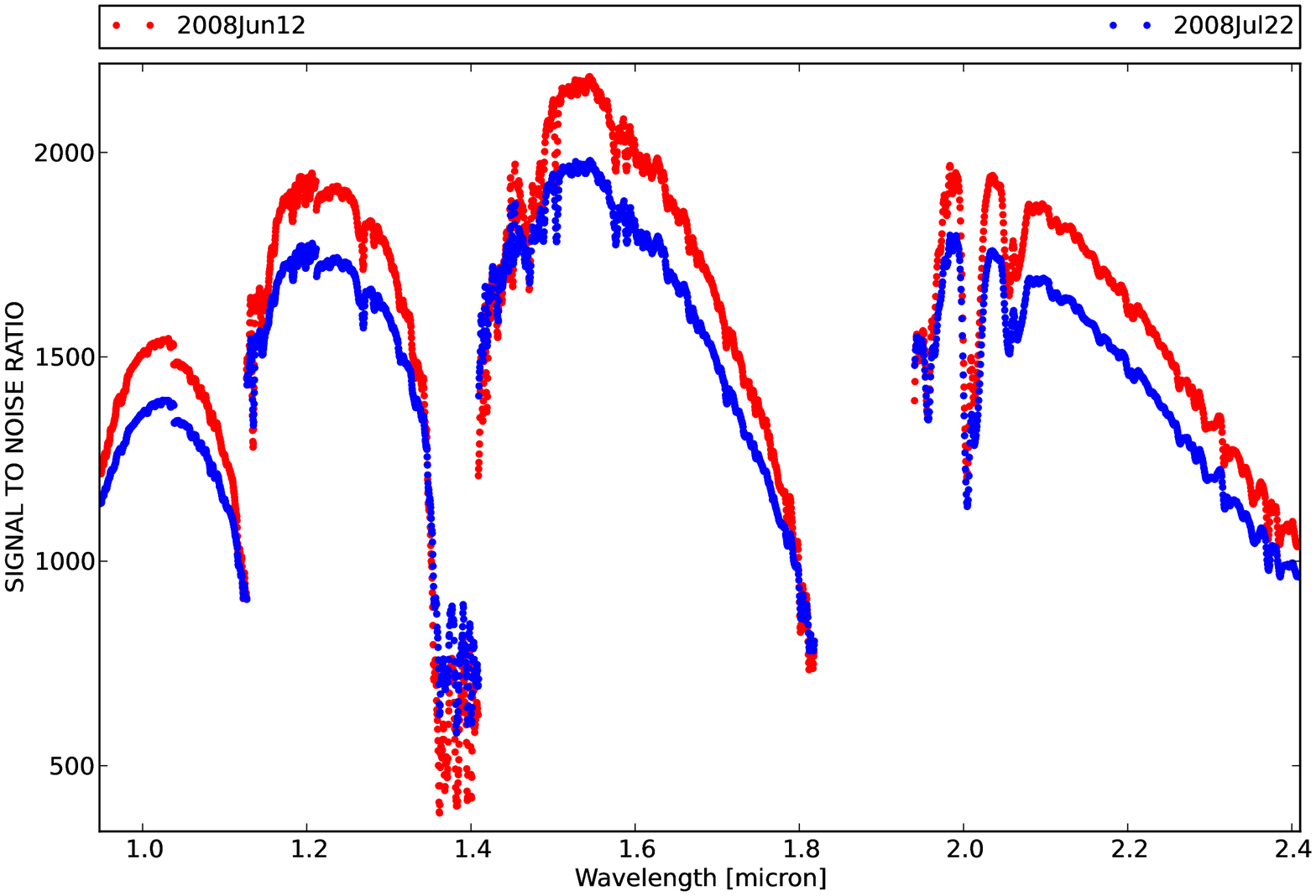}
\caption{Signal-to-noise ratio measured over the totality of the eclipse event for two nights of observation.}
\label{SNR}
\end{figure}

We describe here the observations  obtained on 12 June 2008 and  22 July 2008, using the SpeX 
spectrograph at the NASA Infrared Telescope Facility.
The instrument was configured in the short-wavelength cross-dispersed mode (SXD: 0.8-2.5  $\mu$m), in which five orders of the 
spectrograph covered the spectral range from 0.8 to 2.5 $\mu$m simultaneously, 
with a small gap in wavelength coverage due to a strong telluric absorption feature between 1.81 and 1.86 $\mu$m (Rayner et al. 2003).
We used  a slit of 1$\farcs$6$\times$15$\farcs$, an average integration time of 6 sec and the ABBA nodding sequence. Flat field and argon lamp calibrations 
were obtained before and after the transit.
The observations were scheduled to start two hours before the planetary ingress and to stop two hours after the planetary egress.
Comparing the two nights, 22 July 2008 presents a higher scatter and a gap in the lightcurves between -0.032 and -0.028 of the normalized orbital phase, 
due to a switch-off of the instrument. Moreover its signal-to-noise ratio (SNR) is lower than the SNR obtained over 
the night of 12 (Fig. \ref{SNR}).  An additional transit event has been observed with the same settings on 23 June 2010, but was discarded due to the unrecoverable 
changes of throughput during the planetary ingress.

\subsection{\textit{Data analysis}}
 \label{datanalysis}

We used the SpeX data reduction package, Spextool (Cushing et al. 2004) and we applied the standard calibration procedure, which includes background 
subtraction, flat fielding and wavelength calibration. This procedure yielded sets of 382 and 450 individual stellar spectra for the first and the second night respectively. 
We removed  the outliers by sigma clipping the 
time-series $T_{\lambda}$ at 4$\sigma$  and we replaced the bad pixels with the mean of the surrounding pixels.
We  applied a first-order airmass correction, using the cosine-based approximation provided by David Tholen in the IDL library. We then obtained 
the modified time-series $T'_{\lambda} = T_{\lambda} \times ce^{-w \cdot \textrm{airmass}}$.

To extract the planetary signal from the raw data, we used de-noising techniques followed by amplification of the signal.
We applied the \textit{Model-Correlate-Fit} (MCF) method (Swain et al. 2010, Waldmann et al. 2012) 
to the data, which is useful when the wavelength correlated noise is stronger than the channel-to-channel differential signal. The MCF approach allows the extraction of the exoplanetary 
signal by correlating multiple time-series in the Fourier domain. 
Since here we treat primary transit observations, we  removed the average transit depth
in the J-, H- and K- bands by dividing by the mean flux of the analysed band. The final spectrum is thus a differential transmission spectrum.

A critical step in our analysis is the cleaning of all the non-random temporal changes in the spectro-photometric 
time-series that are not strictly associated 
with the exoplanetary transit (i.e. variations of pixel sensitivity, bias offsets). 
We therefore normalised each spectrum $F_{i}$ ($i$ corresponds to each spectral frame of the dataset)
to remove all the systematic errors correlated in wavelength: 

\begin{equation}
F^*_{i} (\lambda) = \dfrac{F_{i}(\lambda)}{\overline{F_B}} 
\end{equation}

\noindent where $F^*_{i}$ is the normalized spectrum, $F_{i}$ is the flux of each spectrum $i$ as a function of wavelength $\lambda$ and
$\overline{F_B}$ is the mean flux of the analysed band, which ranges from $\lambda_{1}$ = 0.94 $\mu$m to $\lambda_{2}$ = 1.39 $\mu$m for the J-band, 
from $\lambda_{1}$ = 1.41 $\mu$m to $\lambda_{2}$ = 1.81 $\mu$m for the H-band and 
from $\lambda_{1}$ = 1.94 $\mu$m to $\lambda_{2}$ = 2.42 $\mu$m for the K-band:

\begin{equation}
\overline{F_B} = \dfrac{\int_{\lambda_1}^{\lambda_2} \! F_{i}(\lambda) \, \mathrm{d} \lambda}{\lambda_2 - \lambda_1}
\label{mean_band}
\end{equation}

\noindent It is important to stress that the normalisation of the broad-band flux does not erase the wavelength-dependent spectral modulations.

\begin{figure}[f]
\subfigure{\epsfig{file=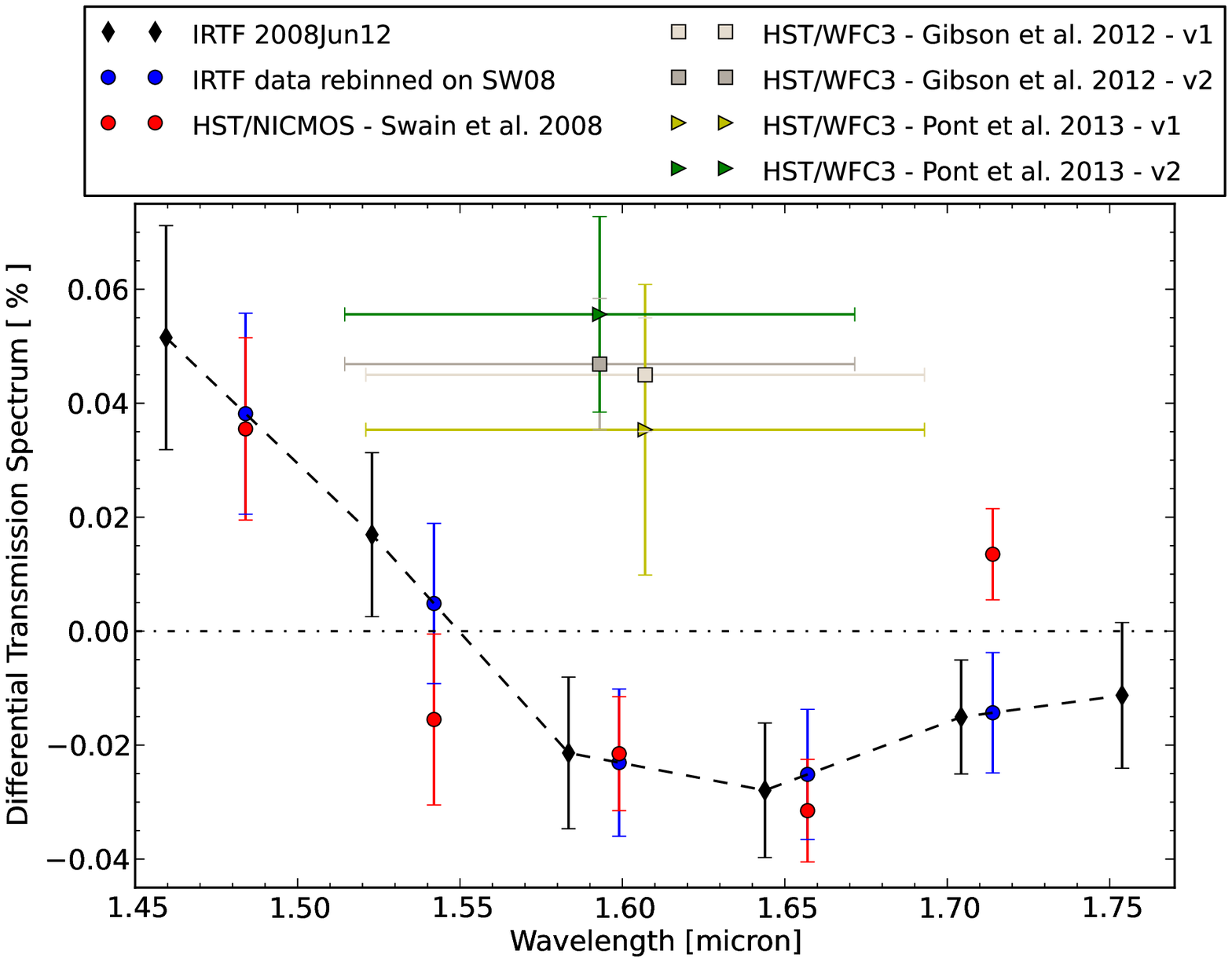,width=0.5\linewidth,clip=, trim=  10 10 30 10}}
~
\subfigure{\epsfig{file=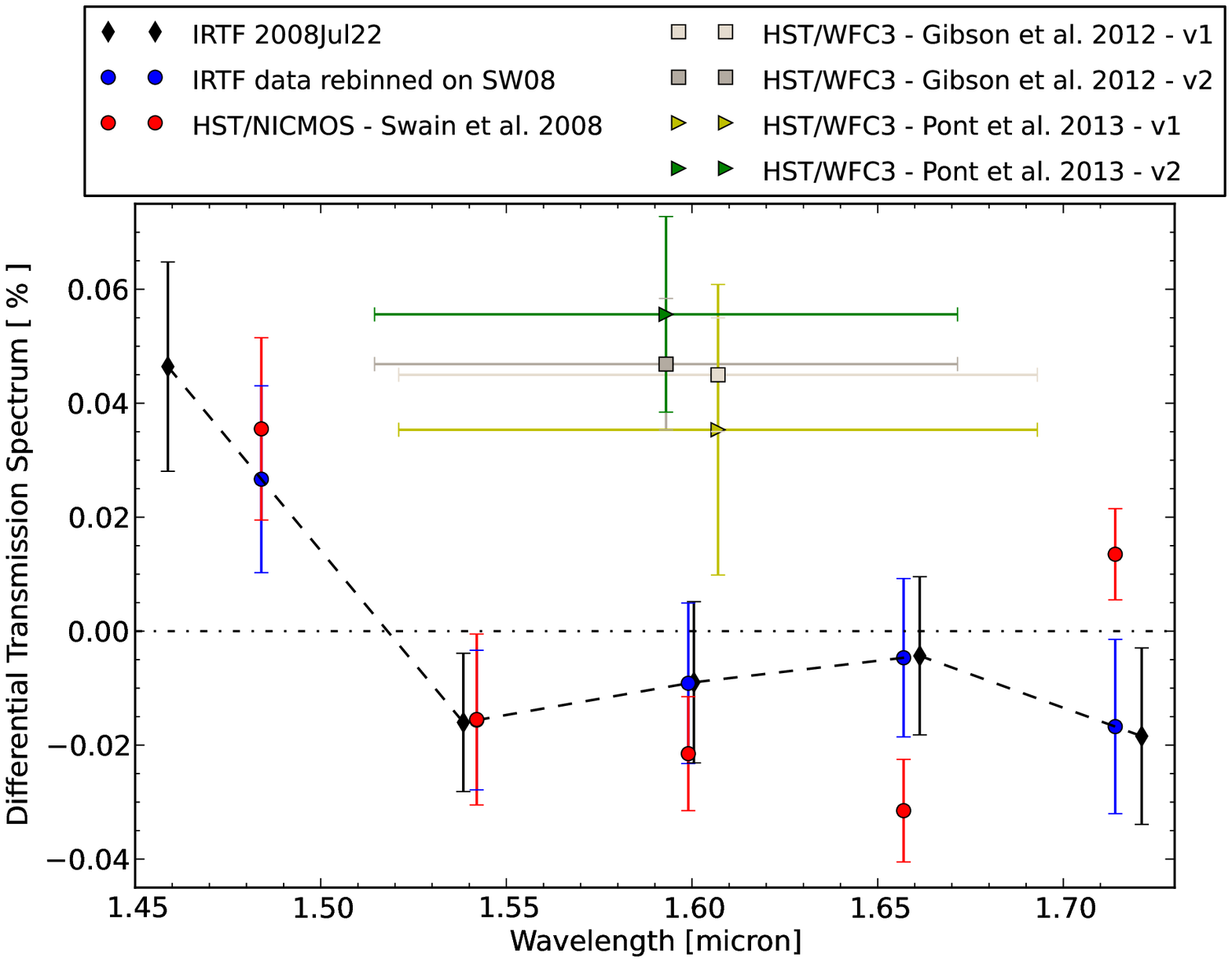,width=0.5\linewidth,clip=,  trim=  10 10 30 10}}
\caption{H-band differential transmission spectrum of the hot Jupiter HD-189733b (black points), with uncertainties of $\pm$ 1$\sigma$.
For comparison we show the SW08 spectrum (red dots) and the IRTF points rebinned on the NICMOS wavelengths (blue dots). 
We also show the non-simultaneous data points recorded with the Hubble/WFC3 in visit one -- v1 and visit two -- v2 (light and dark grey squares, Gibson et al. 2012b, yellow and green triangles, Pont et al. 2013). \textit{Left}: 12 June 2008, \textit{Right}: 22 July 2008.}
\label{Hband}
\end{figure}

\begin{figure}[f]
\subfigure{\epsfig{file=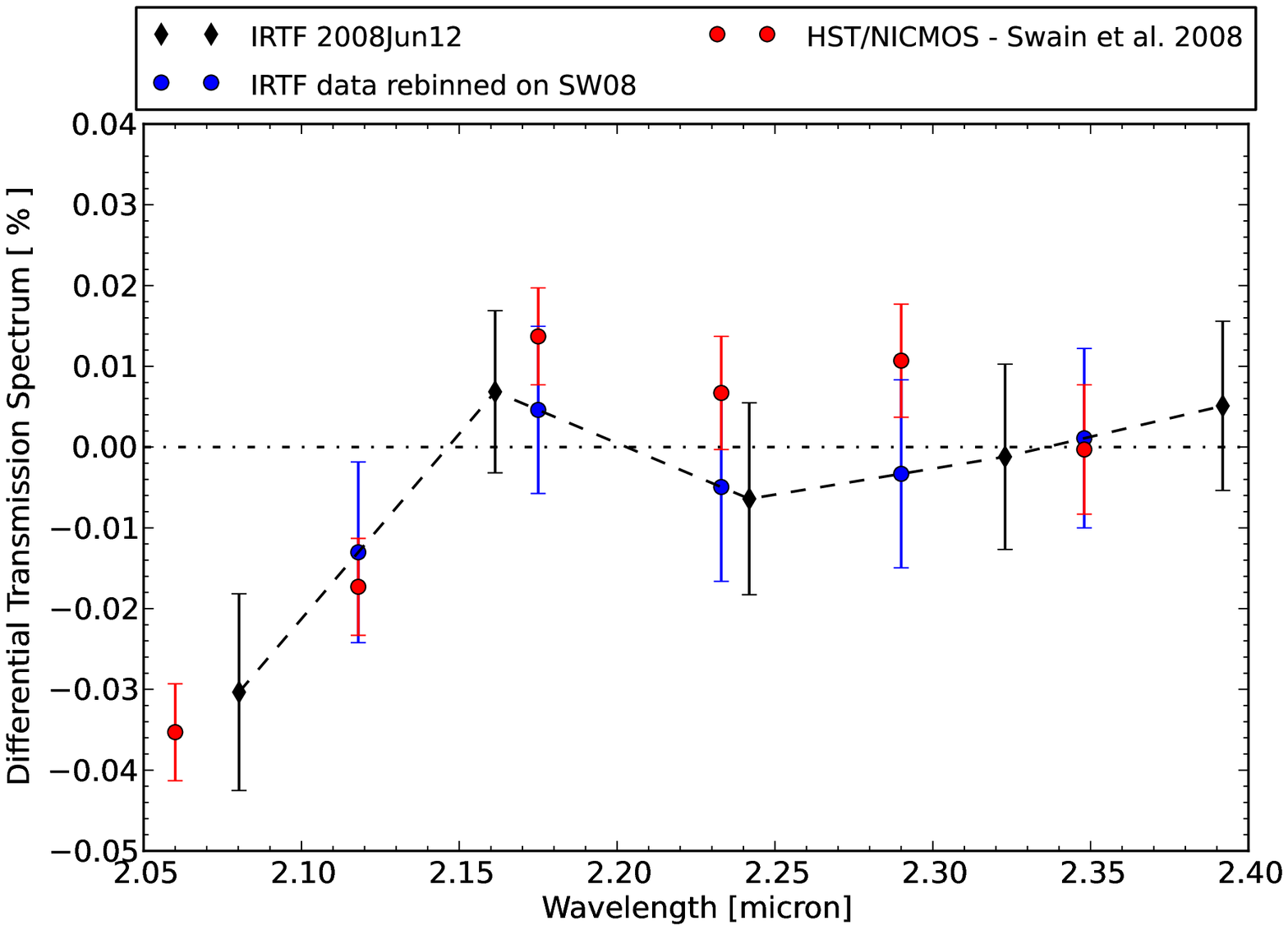,width=0.5\linewidth,clip=, trim=  10 10 30 30}}
~
\subfigure{\epsfig{file=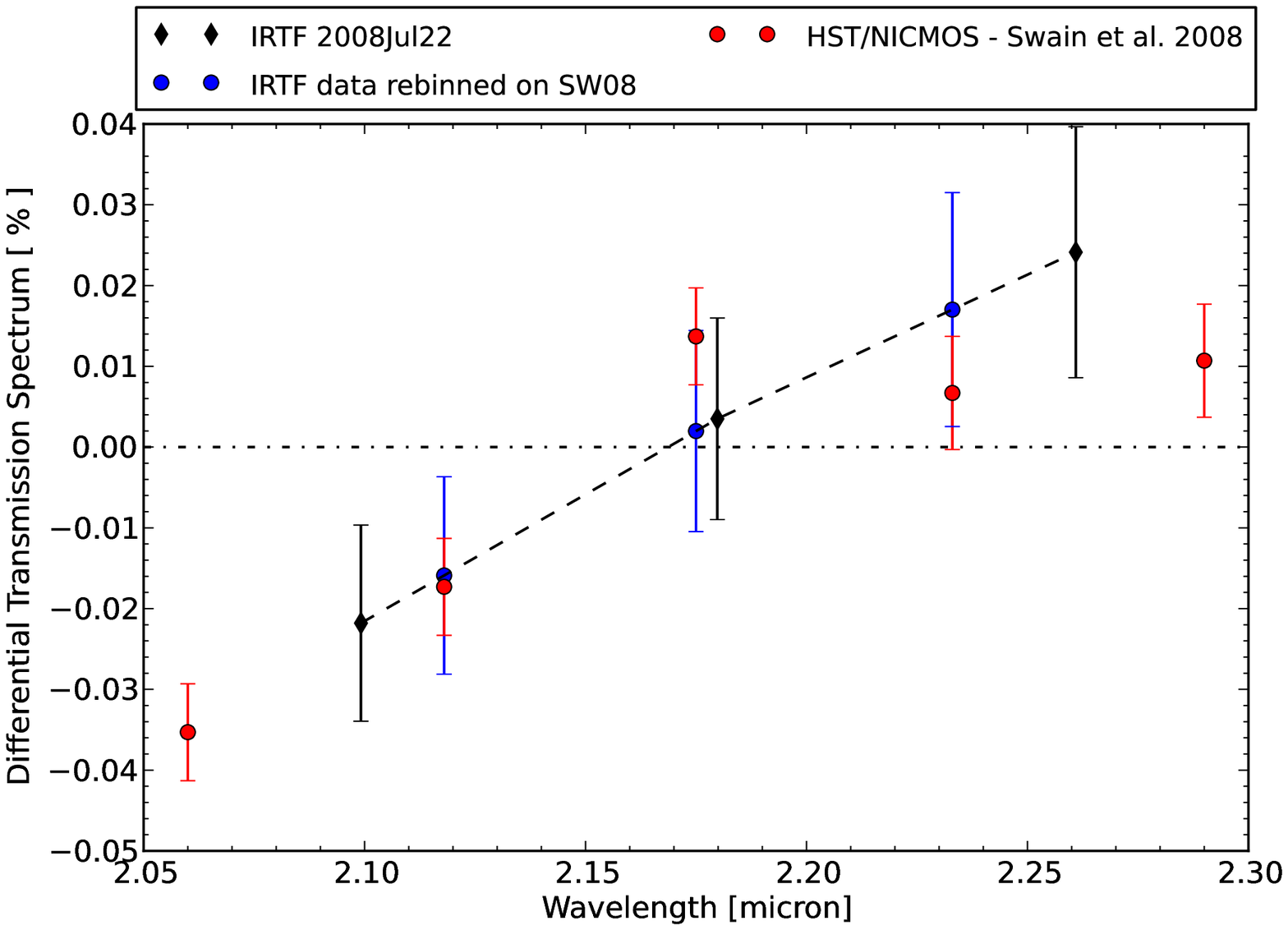,width=0.5\linewidth,clip=,  trim=  10 10 30 30}}
\caption{K-band differential transmission spectrum of the hot Jupiter HD-189733b (black points), with uncertainties of $\pm$ 1$\sigma$.
For comparison we show the SW08 spectrum (red dots) and the IRTF points rebinned on the NICMOS wavelengths (blue dots). 
 \textit{Left}: 12 June 2008, \textit{Right}: 22 July 2008.}
\label{Kband}
\end{figure}

Among the data sets analysed, the one recorded on 12 June 2008 presents fewer systematics and is more uniform in terms of scatter and flux modulations.
In contrast, 22 July 2008 data exhibit a higher scatter, especially during the pre-ingress, 
most likely due to the poorer atmospheric conditions (i.e. probably the presence of a thin layer of cirrus).  
We noticed that for highly scattered time-series, the removal of the mean broad-band flux caused an offset between the A and B frames obtained with the nodding technique. 
These problematic time-series, either at the edges of the bands analysed or  associated with
low SNR (Fig. \ref{SNR}), were discarded
to prevent the addition of systematics.   However, one needs to be cautious about this process:
while a proper correction must be applied to eliminate the systematics, too few spectral channels do not allow us to extract a meaningful signal.
The number of data points and the uncertainties in the final spectra (Fig. \ref{Hband}, \ref{Kband}) are directly related to this critical step. 

\begin{figure}[f]
\epsscale{0.5}
\plotone{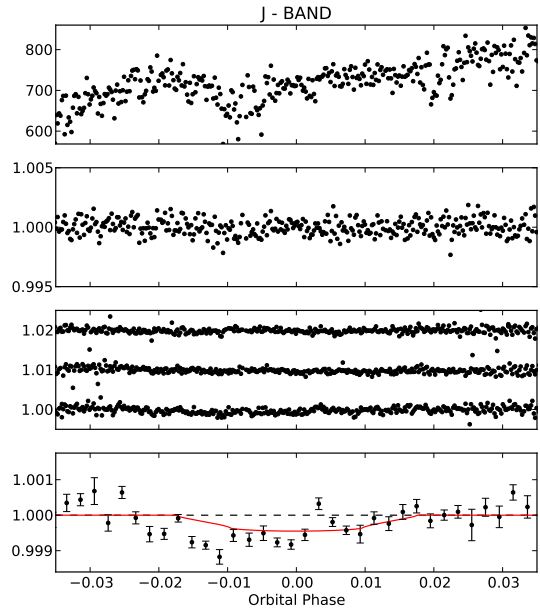}
\caption{From the top: (1) Raw flux for a single pixel-based channel in the J-band, (2) Light curve of a single channel after 
airmass correction and normalization, (3) A sample of three sets of 150 channels correlated together 
in the Fourier domain, (4) Detrended and binned light curve of one of the J-band set with an eclipse model overplotted.}
\label{block_diagram}
\end{figure}

We applied a second normalisation to remove residual fluctuations in the 
time-series, obtaining $F^{**}_{i} (\lambda)$:

\begin{equation}
F^{**}_{i} (\lambda) = \dfrac{F^*_{i} (\lambda)}{\overline{F_X}} 
\end{equation}

\begin{equation}
 \overline{F_X} = \dfrac{\sum_{a}^{b} F^*_{i}(\lambda)}{X} 
\end{equation}

\begin{equation}
a = (\lambda_1+\kappa\cdot X),\  b = (\lambda_1+(\kappa+1)\cdot X)
\end{equation}

\noindent where $X$ is the number of spectral channels we combined  and $\overline{F_X}$ is the mean flux of the $X$ channels. $\kappa$ represents a set of $X$ channels. 
$X$ should be optimised depending on the quality of the dataset. We found that usually $\sim$ 100 to 150 was a good compromise between the SNR requirements and 
the spectral resolution. 
The second panel in Fig. \ref{block_diagram} shows a lightcurve for a single channel after all the corrections.

After the pre-cleaning process, each lightcurve $T''_\lambda$ is still too noisy to show the transit. 
We extracted the signal common to all time-series by Fourier-transforming all the time-series
in the spectral band considered and by stacking together the $X$-transformed spectral channels (Fig. \ref{block_diagram}, third panel).
In effect, we are taking the geometric mean of \textit{X} channels. As a result, the spectral resolution is reduced. 
Finally, we take the inverse Fourier transform to convert the data back into the time domain:

\begin{equation}\
T^*_\lambda(t) =  \mathscr{F}^{-1}( \prod^{X}_{\lambda = 1} \mathscr{F}[T''_\lambda(t)] )^{1/X}
\label{selfcohe}
\end{equation}

\noindent where $\mathscr{F}$ denotes the Fourier transform and $\mathscr{F}^{-1}$ its inverse (see details in appendix \ref{appendix}).

We fitted each of the final lightcurves with two components: a residual baseline curve (characterised by a second order polynomial function) 
fitted to the out-of-transit and a transit lightcurve model (Mandel $\&$ Agol 2002) 
with the transit depth being the only free parameter (Fig. \ref{block_diagram}, bottom panel). We used the limb-darkening coefficients by Claret (2000).
The transit depths $\delta_{\kappa}$ of these final lightcurves are the differential transmission spectra.
We repeated the same procedure for J-, H- and K- bands for both nights.

To calculate the error bars, we considered both the standard deviation of the model subtracted residual, $\sigma_{SCAT}$, and the standard deviation of the
transit depth derived from changing baseline fits, $\sigma_{FIT}$. The final standard deviation $\sigma$ is given by the quadrature of the two previous terms:
$\sigma = \sqrt{\sigma_{SCAT}^2 + \sigma_{FIT}^2}$.

\begin{figure}[f]
\epsscale{0.8}
\plotone{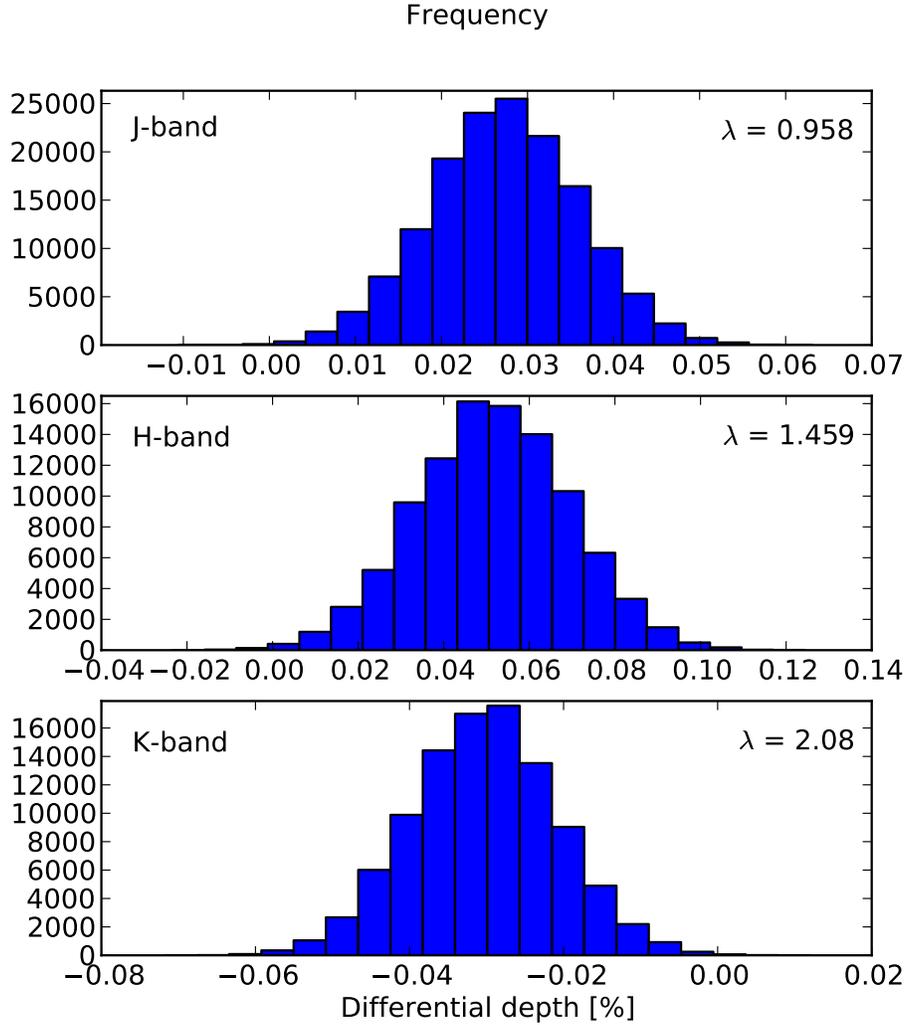}
\caption{Histograms of the depths retrieved with the MCMC (12 June 2008). From the top: lightcurve at 0.958  $\mu$m (J-band), lightcurve at 1.459 $\mu$m (H-band) 
 and lightcurve at 2.08 $\mu$m (K-band). Note that the standard deviations measured with the MCMC are smaller than those computed with the MCF technique.}
\label{histogram}
\end{figure}

To test the robustness of our results, we used a Markov-chain Monte Carlo method (MCMC) to re-fit the lightcurves.
We did two MCMC fits, the first one to model the residual baseline curve and the second one to fit the transit lightcurves.
The posteriors of the first fit were used to create a covariance matrix. From the same covariance matrix we generated a multivariate prior for
the second fit. The retrieved depths and standard deviations are consistent with our previous analysis (Fig. \ref{histogram}).

\begin{figure}[f]
\centering
\epsscale{0.9}
\epsfig{file=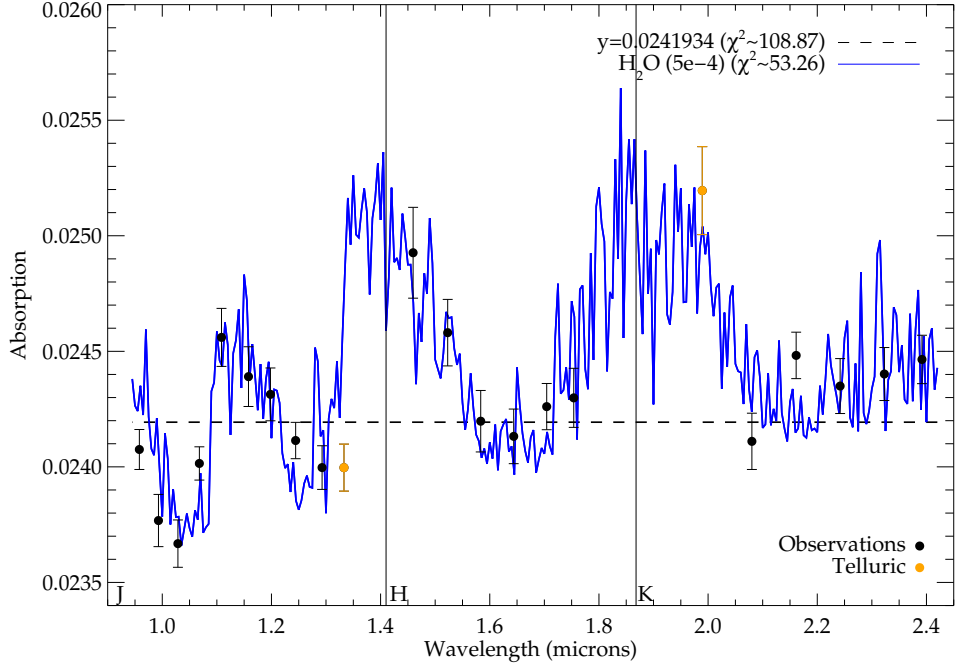,width=0.8\linewidth,clip=} \\
\caption{0.94 - 2.42 $\mu$m IRTF transmission spectrum compared to a simulated spectrum for water with a mixing ratio of $5 \cdot 10^{-4}$, assuming an isothermal 
atmosphere at $T\sim1500$K. The $\chi^2$ value for this fit is also given, where the data values known to suffer from telluric contamination (marked `Telluric' in the figure legend) 
were excluded for this calculation. For reference, a straight line with a value equal to the mean of all of the data points ($y=0.0241934$) is also shown, with the associated $\chi^2$ value.
For comparison, when optimising the spectrum to the flat line we obtain a $\chi^2$ =74.05. }
\label{water}
\end{figure}

\begin{figure}[f]
\centering
\begin{tabular}{c}
\epsfig{file=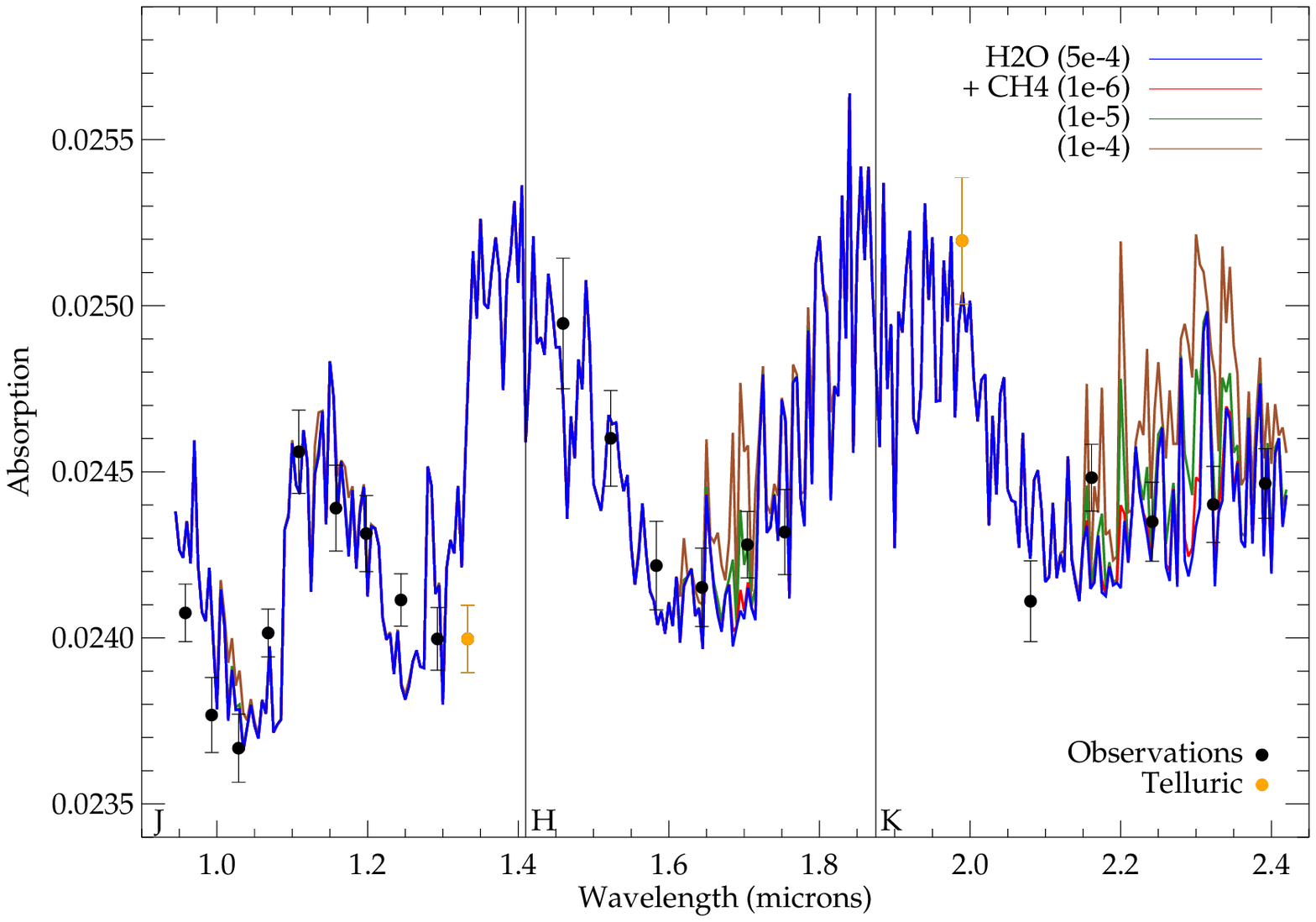,width=0.5\linewidth,clip=} \\
\epsfig{file=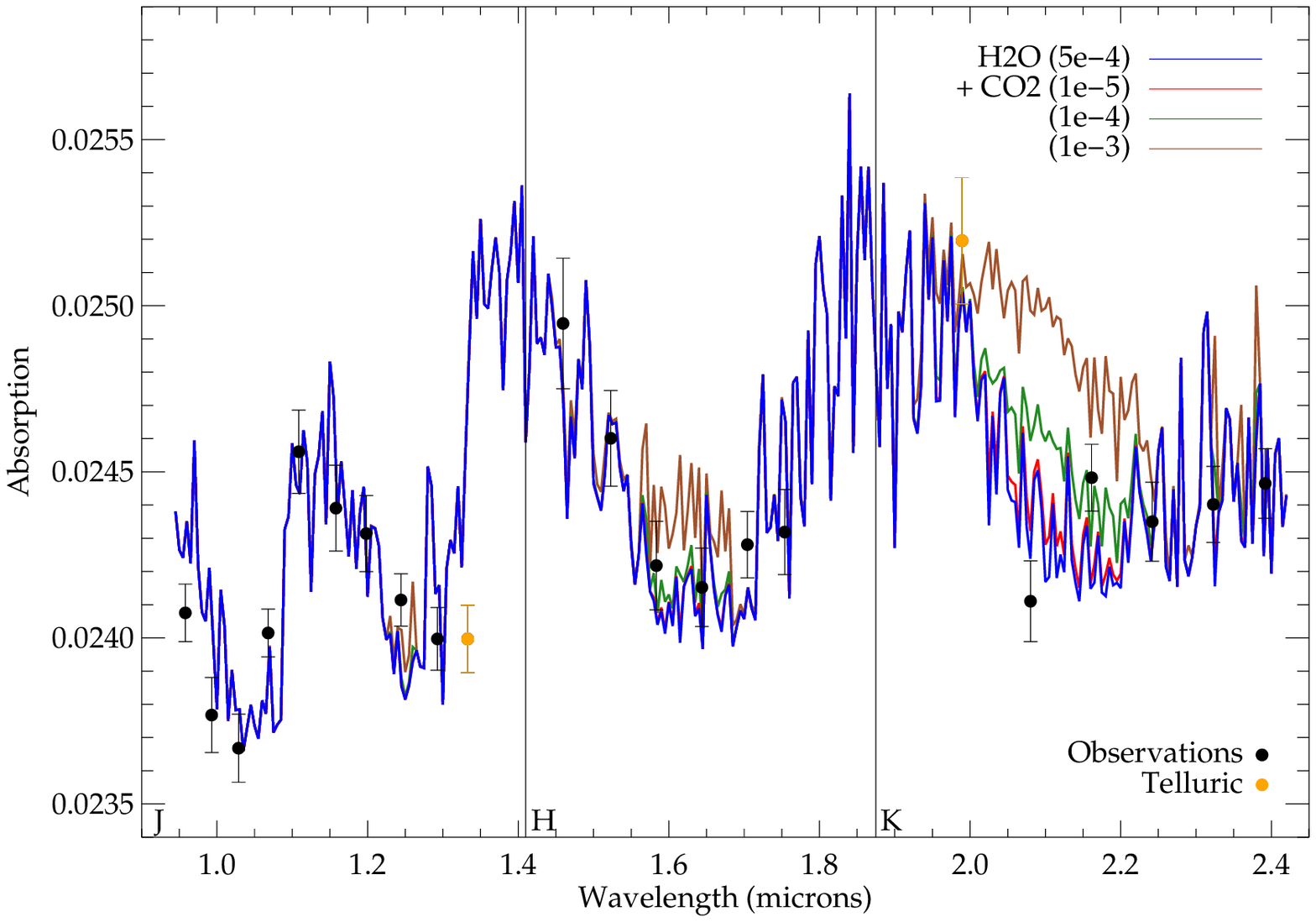,width=0.5\linewidth,clip=} \\
\epsfig{file=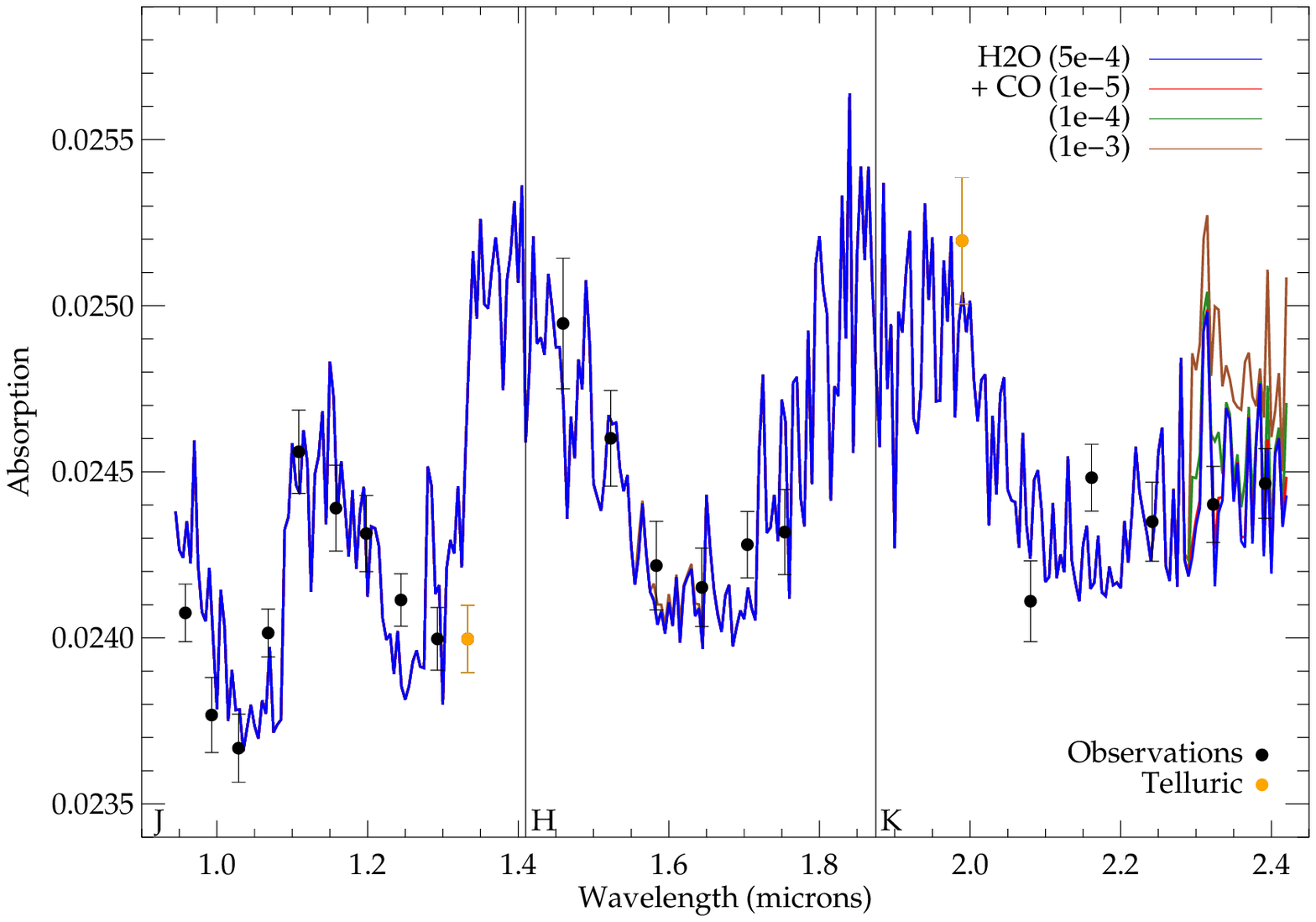,width=0.5\linewidth,clip=} \\
\end{tabular}
\caption{0.94 - 2.42  $\mu$m IRTF transmission spectrum compared to different families of simulated spectra. Each family includes water with mixing ratio $5 \cdot 10^{-4}$
plus varying abundances of methane (top), carbon dioxide (center) and carbon monoxide (bottom), assuming an isothermal atmosphere at $T\sim1500$K.
Water alone can explain well the features of our IRTF transmission spectrum; the data are not sensitive enough to give constraints on the presence of other molecules.}
\label{models}
\end{figure}

Additionally, combining the spectra obtained for the 12 June 2008 night we obtained a transmission spectrum of 
HD-189733b that covers the wavelength range from 0.94-2.4 $\mu$m.
As the wavelength variation shown by the result is indicative of the presence of spectral features, we proceeded to analyse this aspect.
We thus compared the final transmission spectrum with synthetic models of the planetary atmosphere (see Sect. \ref{results}, Fig. \ref{water}, \ref{models}). 
Note that this comparison is possible as all the J-,H- and K- band spectra were recorded simultaneously.

\section{Models} 
 \label{model section}
The transmission spectra were modelled using a line-by-line radiative transfer code as described in Tinetti et al. (2007), Tinetti et al. (2012) and Hollis et al. (2013). For all models, an isothermal atmospheric profile at $T\sim1500K$, probing down to a pressure of 10 bar, was used. The atmosphere was taken to be cloud-free initially, but Rayleigh scattering due to molecular hydrogen (Liou 2002) was included, determined assuming a bulk composition of $85\%$ molecular hydrogen and $15\%$ atomic helium, with particle sizes and refractive indices from Vardya (1962) and Allen (2000). Thus the mean molecular weight of the atmosphere was taken to be $2.3\,g\,mol^{-1}$, giving a pressure scale height of approximately $230\,km$ for example at a temperature of $1500\,K$. H$_{2}$ - H$_{2}$ collision-induced absorption (hereafter: CIA - Borysow et al. 1997) was also included - this continuum (plus Rayleigh scattering at the shorter wavelengths) provides an effective absorption floor, on top of which was added absorption from H$_{2}$O, CH$_{4}$, CO$_{2}$ and CO, these being the molecules previously detected on the dayside (Swain et al. 2009, Madhusudhan $\&$ Seager 2009, Lee et al. 2012, Line et al. 2012). The mixing ratios (abundances) of these molecules were varied in order to determine the best composition, i.e. to provide an adequate fit to the observations. 

For most of the molecules considered, theoretical line lists from the \textit{Exomol} project\footnote{http://www.exomol.com} (Tennyson et al. 2012) were used. For the H$_{2}$O molecule, absorption cross-sections were obtained from Barber et al. (2006), which contributes absorption features in the 0.94 - 2.42 $\mu$m range in particular. The CH$_{4}$ opacity was modelled using a preliminary version
of a new variational line list at high temperature (T=1000K; Yurchenko et al., in prep.). For the CO and CO$_{2}$ molecules, the \textsc{hitemp} line list (Rothman et al. 2010) was used. 

The cloud opacity was calculated using the Mie-Lorenz scattering approximation given by Liou (2002), with the cloud particle size taken to be single-valued rather than a distribution around some mean size, for simplicity and to avoid further model degeneracies where possible. Even with this simplified approach though, it can be seen that it is possible to fit both data with a Mie-like signature at small wavelengths with those data exhibiting spectral features due to molecular absorption at longer wavelengths. Thus multi-band observations that have previously been claimed to be mutually inconsistent can potentially be drawn together into a single model framework. A more rigorous approach including more physically-motivated cloud properties (e.g. using various cloud particle size distributions, and appropriate refractive indices) would allow further tuning of such models for better fits to the observations. However, since the different data were measured at different epochs with different instruments, and are further subject to varying systematic offsets and stellar activity levels, combining and comparing such datasets must always be done with the utmost caution, with any inferences also subject to these caveats.

\section{Results}
\label{results}
\begin{figure}[f]
\begin{center}
 \epsscale{1.0}
 \plotone{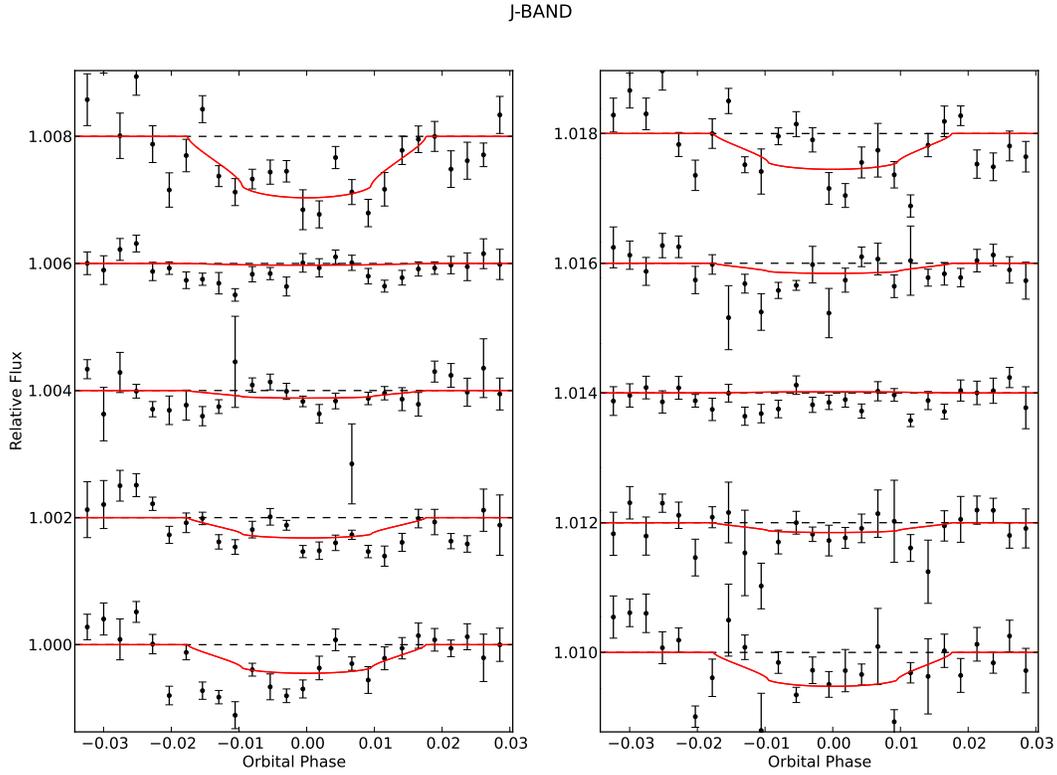}
\end{center}
\caption{The 12 June 2008 detrended lightcurves for the J-band showing (in red) the fitted differential lightcurve model.
The uncertainties of the individual measurements are $\pm$ 1$\sigma$ deviation. The wavelengths 
are increasing from bottom to top and left to right, and the light curves have been shifted for clarity.}
\label{J_lightcurves}
\end{figure}

\begin{figure}[f]
\begin{center}
 \epsscale{1.0}
 \plotone{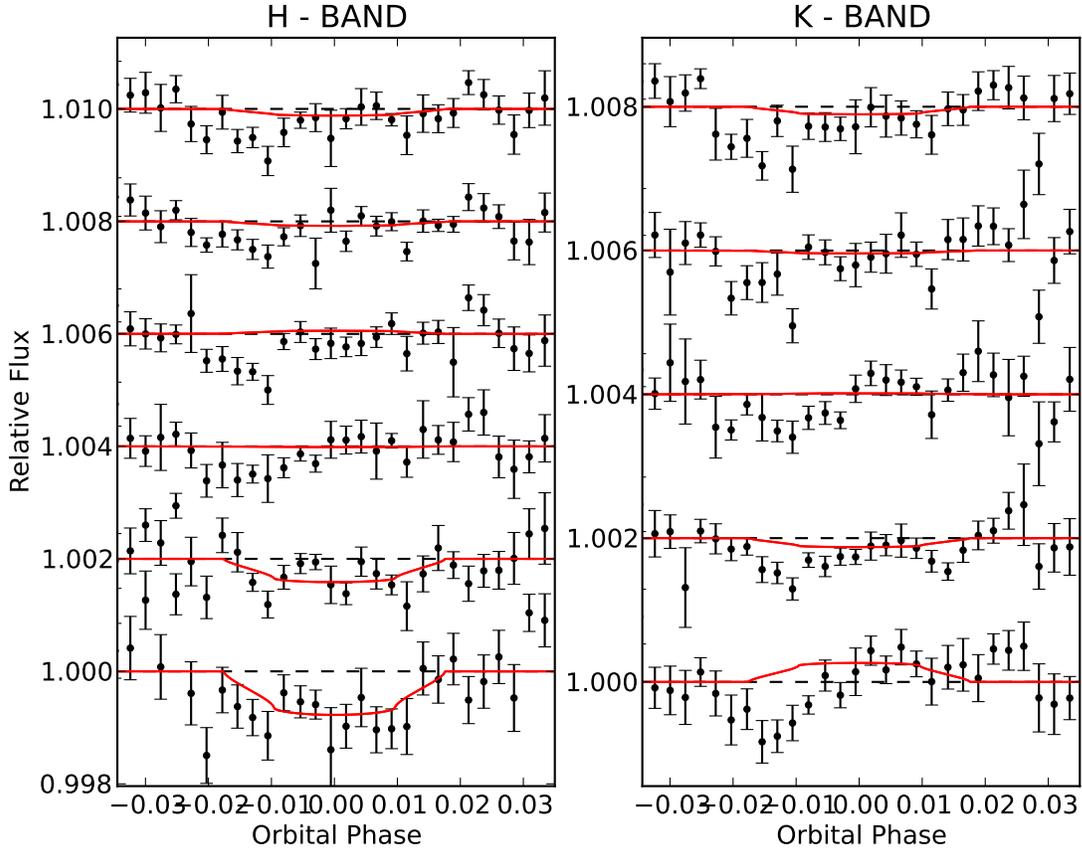}
\end{center}
\caption{The 12 June 2008 detrended lightcurves for the H-band (left) and the K-band (right) showing in red the fitted differential lightcurve model.
The uncertainties of the individual measurements are $\pm$ 1$\sigma$ deviation. For both bands, all the lightcurves are shifted for clarity and the wavelength 
is increasing from the bottom to the top.}
\label{lightcurves}
\end{figure}

Using the method described in the previous section, we analysed the J-, H- and K-band data for both nights.
By combining sets of 150 channels we
extracted the differential depths  as a function of $\lambda$ (Table \ref{depth}). We show in Figs. \ref{J_lightcurves} and \ref{lightcurves} the  lightcurves for 
the first night, J- and H-,  K- respectively.
Notice that this approach does not provide an accurate absolute calibration, but rather a relative measurement. 
Uncertainties much higher than the photon-noise  and the residual correlation are typical of ground-based observations and
depend on the quality of the data.
The uncertainties in our data are, on average for the first night, three times larger than photon noise in the H-band and two times larger in the J- and K-bands. 
For the second night the uncertainties  for the H- and K- bands are three times larger than the photon noise. The J-band on the second night showed a lower signal level.\\

\begin{figure}[f]
 \begin{center}
  \epsscale{1.}
  \plotone{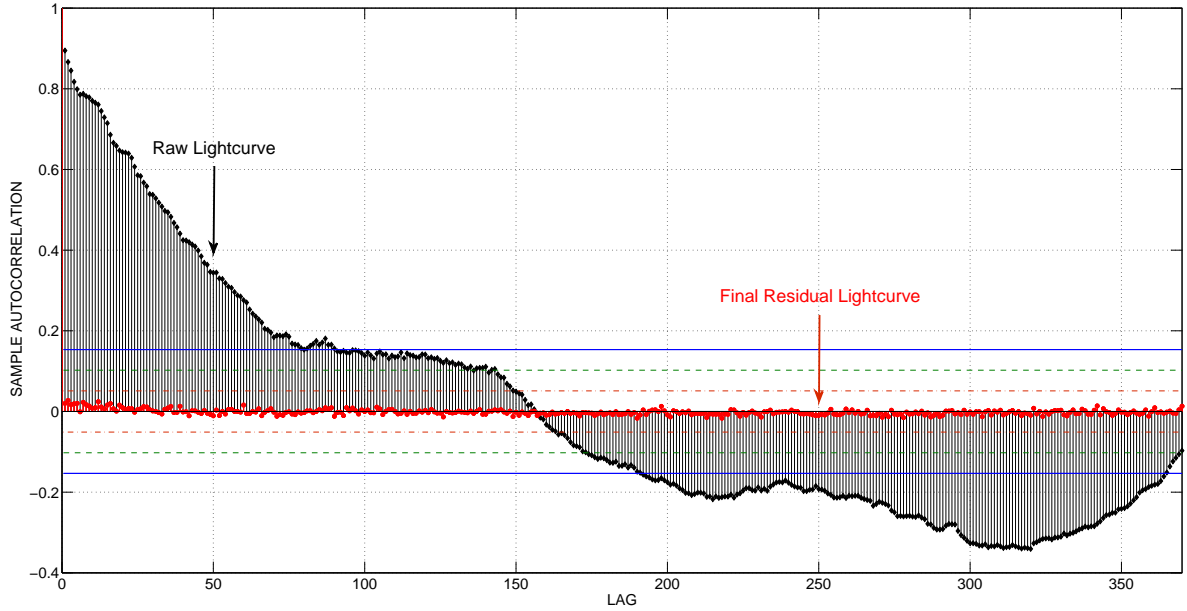}  
 \end{center}
\caption{Showing the auto-correlation function of one raw lightcurve (black diamonds) and one final lightcurve obtained after our data analysis (red points) for the H-band 
 of the 12 June 2008 night.
The 3$\sigma$, 2$\sigma$, 1$\sigma$ confidence limits that the data are normally distributed are plotted in blue (solid line), green (dashed line) and red (dash-dot line).
All the lags are within 1$\sigma$ limit showing that the correlations were efficiently removed, diminishing the auto-correlative noise at most frequencies.}
\label{autocorr}
\end{figure}

To test the statistical significance of our results we estimated the auto-correlation function (ACF) for our time-series before and after
the data reduction (Fig. \ref{autocorr}). Given a discrete signal $S_n$, the ACF($\tau$) is the cross-correlation of the signal $S_n$ with itself, 
at lag $\tau$:

\begin{equation}
 ACF (\tau) = \sum_{n} \! S_n \  \overline{S}_{n-\tau} 
\end{equation}

\noindent where $\overline{S}$ denotes the complex conjugate.
For wide sense stationary white noise, the ACF will show a peak at $\tau$ = 0 
with all other lags within the one sigma bounds of white noise. 
Values outside these bounds indicate a significant autocorrelation (in this case in time) of the signal at the respective lag.
In Fig. \ref{autocorr} we present the ACF over 370 lags for one raw lightcurve and for the
residual to the final lightcurve. The raw data  present a strong correlation. By contrast, the autocorrelation for final residuals is all  within a 1$\sigma$ confidence 
level  and their amplitudes are negligible. This indicates that correlations in time were efficiently removed and that our results are normally distributed and 
auto-correlative noise is decreased at most frequencies.
At lower lags, some residual correlation persists but these systematics are at frequencies higher than the time scale of 
the transit event of HD-189733b, hence the signal should not be directly affected.\\

\begin{figure}[f]
\begin{center}
 \epsscale{1.}
 \plotone{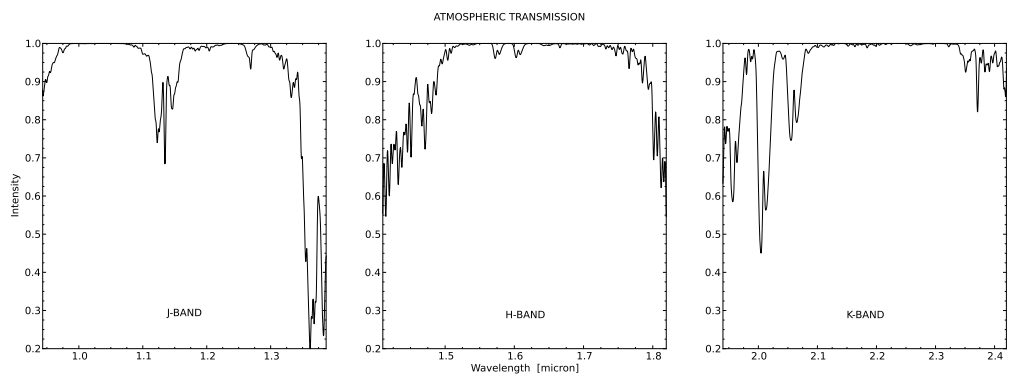}
 \end{center}
\caption{Atmospheric transmission on Mauna Kea over the three analysed bands. These data, produced using the program IRTRANS4, 
were obtained from the UKIRT web pages (airmass = 1.0, H$_{2}$O = 1.2 mm, R = 3000).}
\label{transmission}
\end{figure}

%
Given the good atmospheric transmission window from 0.94 -1.30 $\mu$m (Fig. \ref{transmission}) and the good quality of the 
12 June 2008 data, we compared the J-band spectrum to synthetic models 
that included H$_{2}$O, CH$_{4}$, CO$_{2}$ and CO, these being the molecules detected in the dayside of the planet (Fig. \ref{models}). 
We tested one molecule at a time followed by combinations of these.
The best fit for the J-band was obtained with water with a mixing ratio of 5$\cdot 10^{-4}$ at a temperature of T $\sim$ 1500K. 
We then compared the best model achieved with the whole 0.94 - 2.42 $\mu$m spectral range.
Note that our J-, H- and K-band spectra are relative measurements, hence their absolute level is unconstrained.
It is therefore possible to renormalise their offsets. Thus, keeping our J-band spectrum fixed, we matched the H- and K-band spectra to the model, 
finding an excellent agreement.
The best fit was given by the model that includes water with a mixing ratio of 5$\cdot$10$^{-4}$ at a temperature that ranges between T$\sim$1000 - 1500 K.
We also tested the possibility of a flat spectrum by fitting a straight line to the data, however this gave a poor fit as confirmed by the $\chi^2$
(see Fig. \ref{water}).

\section{Discussion}
While the spectra obtained during the two nights are consistent with each other within the error bars, the spectra
 extracted on the first night are of superior quality.
Among the hurdles encountered, we can list the small gap in the flux during the orbital phase, the instrumental systematics 
(e.g. variations of the pixel sensitivity), the residual telluric contamination, 
and the presence of high altitude clouds during the second night. 
All these effects may influence our results at different levels, but by applying the MCF method to the IRTF data
we have been able to extract statistically significant planetary spectra. The ACF test in fact proves that our results 
are normally distributed and auto-correlative noise is decreased at most frequencies.
Nonetheless these results could be largely improved by combining together a large number of good nights.
As demonstrated by Waldmann et al. (2012), we can potentially amplify our signal by a factor $\sim N$
by combining  $N$ multiple good observations.
In the work presented here two of the nights could be used including one of excellent quality, one was unusable.
We decided to flag the measurements at 1.35 $\mu$m and 2 $\mu$m (marked in Figs. \ref{water}, \ref{models} in yellow) as less reliable. 
At these wavelengths, the telluric contamination is very high and it might affect the measurements (Fig. \ref{transmission}).
In addition, the data point at 1.35 $\mu$m includes fewer spectral channels than our standard procedure, as it lies at the end of the band.

\subsection{\textit{Removal of telluric contamination in the data}}
\label{contamination}
\begin{figure}[f]
 \epsscale{0.6}
\plotone{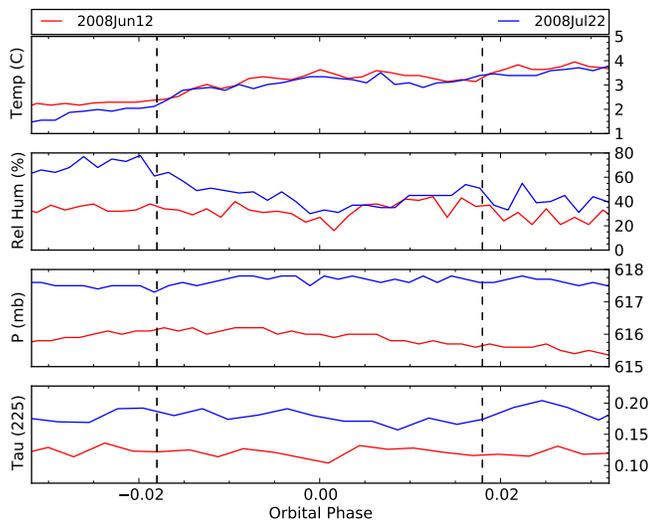}
\caption{Weather over the nights of 12 June 2008 (red) and 22 July 2008 (blue), from the CFHT weather station. Top to bottom: temperature ($\degr$C), 
relative humidity ($\%$), pressure (mb) and the optical depth, tau (225 GHz) from CSO. The dashed vertical lines mark the transit event duration.}
\label{weather}
\end{figure}

The entire MCF method is an efficient procedure to remove, through several steps, the telluric contamination contained in the raw data.
The raw data contain both the signal and the noise components.
These components are difficult to disentangle in the time-domain, but can be described by different sets of frequencies in the Fourier domain.
This is why,  when we combine together $\sim X$ Fourier-transformed lightcurves, noise such as the telluric contamination is weakened, while 
the transit signal is strengthened. We refer to Waldmann et al. (2012) for a detailed explanation.
To quantify the residual contamination by systematics we applied the procedure described in Sect. \ref{datanalysis} only to the out-of-transit lightcurve, 
i.e., removing the eclipse signal.
We find that the amplitude of the systematic noise and the residual telluric component is within the uncertainties of the planetary signal over the whole spectrum.
The effectiveness of the MCF was tested with the ACF. \\
Additionally, we report the weather conditions over the two nights. Fig. \ref{weather} shows the temperature, relative humidity and pressure readings of the
Canada-France-Hawaii Telescope (CFHT)\footnote{http://mkwc.ifa.hawaii.edu/archive} weather station as well as the atmospheric opacity
at 225 GHz by the Caltech Submillimeter Observatory (CSO)\footnote{http://ulu.submm.caltech.edu}. No significant correlations between
these parameters and the expected transit shape was found. 
We show in Fig. \ref{transmission} the atmospheric trasmission on Mauna Kea, obtained in similar conditions to our nights 
(UKIRT\footnote{http://www.jach.hawaii.edu/UKIRT/astronomy/utils/atmos-index.html} web pages).
The intensity of the transmission is on average $I = 1$, with some absorption features around 1.125 $\mu$m, 1.41 $\mu$m, and 2.06 $\mu$m whose effect reflects directly on 
the error bars of the data points evaluated at the same  wavelength. Those error bars are in fact the largest in the J-, H- and K- band respectively.
The strongest telluric absorptions  are reported around 1.355 $\mu$m and 2.0 $\mu$m; hence the decision to flag the two measurements at those wavelengths
as less reliable.

\subsection{\textit{MCF sensitivity validation}}
\begin{figure}[f]
 \begin{center}
 \epsscale{0.7}
 \plotone{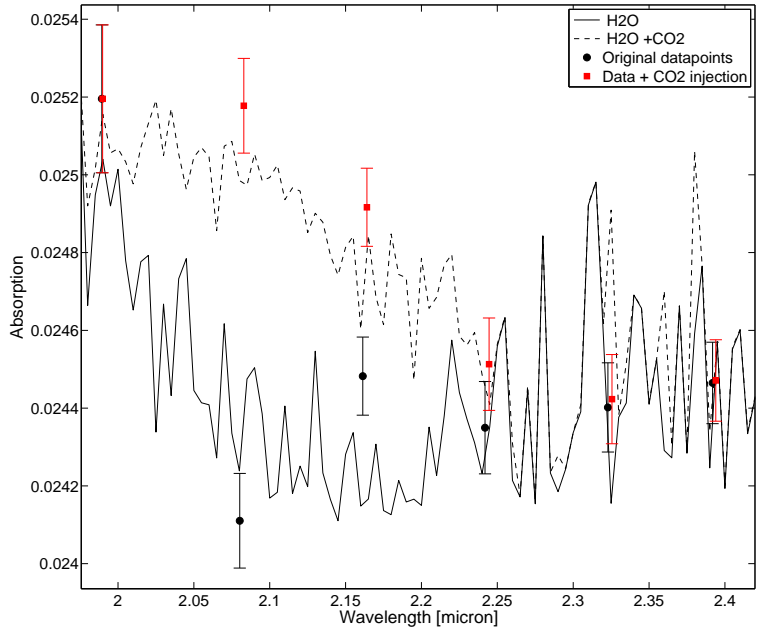}
 \end{center}
\caption{K-band spectrum (red squares) retrieved from our pipeline after the injection into the raw data of a CO$_2$ signal with a mixing ratio of 10$^{-3}$. 
The spectrum is consistent with a model of water and carbon dioxide (dotted line), with mixing ratios respectively of $5\cdot 10^{-4}$ and 10$^{-3}$. 
The original spectrum (black dots) and the fitted water model (continuous line) are shown for reference.}
\label{K_injection}
\end{figure}

To test the robustness of our technique we injected a synthetic planetary spectrum into our data and we proceeded with the MFC process as detailed in \ref{datanalysis}.
For the simulation we chose CO$_2$ as it is an abundant molecule in the Earth's atmosphere. 
We used a mixing ratio of 10$^{-3}$ and an isothermal atmosphere temperature of  T$\sim$1500K. 
The MCF process retrieved the correct amplitude of the artificially injected signal within the errors. Note that the final spectrum is given by a combination of the 
observed one, containing H$_{2}$O,
and the injected one, containing CO$_2$ (see Fig. \ref{K_injection}). 

\subsection{\textit{Comparison with previous observations}}
\label{previousres}
\begin{figure}[f]
\epsscale{0.8}
\plotone{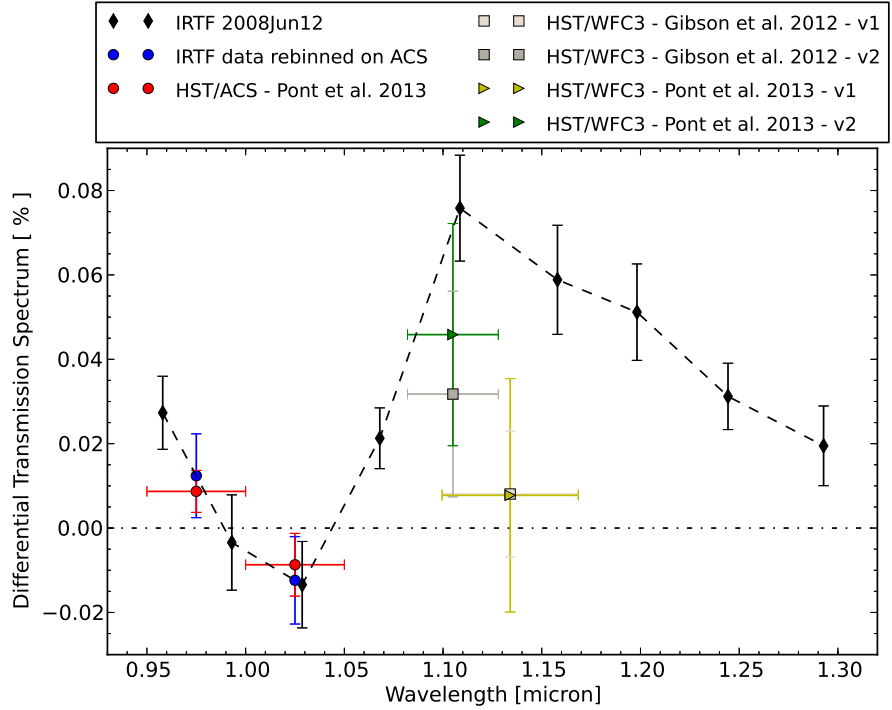}
\caption{12 June 2008 J-band differential transmission spectrum of HD-189733b with uncertainties of $\pm$ 1$\sigma$.
For comparison we show the two ACS bandpass points (red dots, Pont et al. 2013) and the IRTF points rebinned on the ACS wavelengths (blue dots). We also show the non-simultaneous data points recorded with the Hubble/WFC3 in visit one -- v1 and visit two -- v2 (light and dark grey squares, Gibson et al. 2012b, yellow and green triangles, Pont et al. 2013). }
\label{Jband}
\end{figure}

We compared our J-band results with the Hubble/ACS spectrum (Pont et al. 2013) as there is a marginal overlap with the two passband points in the  0.95 and 1.05  $\mu$m range.
We find that the results are consistent at the 1-$\sigma$ level (Fig. \ref{Jband}).
The comparison with the two photometric points recorded with HST/WFC3 (Gibson et al. 2012b, Pont et al. 2013) is less obvious, as these data points are recorded at different epochs
and the star is relative active. Nevertheless we find an agreement within the error bars.

The H- and K-band spectra have been compared to the transmission spectrum of HD-189733b observed by Swain et al. 2008 (hereafter SW08), as it probes the same spectral range.
If we rebin our measurements to the SW08 spectral resolution we find that 
our IRTF differential transmission spectra are consistent  with SW08 at the 2-$\sigma$ level in the H-band 
(Fig. \ref{Hband}) and at the 1-$\sigma$ level in the K-band (Fig. \ref{Kband}). All data are plotted with uncertainties of $\pm$ 1$\sigma$ and  centred on the median wavelength of each set $\kappa$ of channels.\\
Two re-analyses of SW08 were made by Gibson et al. (2012a) and Waldmann et al. (2013). Our data are consistent with both re-analyses
at the 1-$\sigma$ level. Note that our error bars are smaller that the ones estimated by Gibson et al. (2012a) but slightly larger than the ones by Waldmann et al. (2013).
A detailed comparison between Gibson et al. (2012a) and SW08 has been discussed in Waldmann et al. (2013).\\
Our IRTF results for H-band are also consistent with an independent prediction for
the spectral modulation expected by extrapolating models fit to the NICMOS
data (Swain, Line $\&$ Deroo, 2014).
 
Given that the IRTF ground-based measurements and the Hubble space-based measurements have completely 
different sources of systematic errors, the consistency among ground and space observations is quite remarkable.

\subsection{\textit{Data interpretation}}

As mentioned in Sect. \ref{results} we used the 12 June 2008 J-band spectrum and simulated models to calibrate the H- and K- band spectra recorded during the same night.
The models generated included the contributions of H$_{2}$O, CH$_{4}$, CO$_{2}$ and CO. The best fit to the data was given by the water opacity, 
with a retrieved abundance somewhere in the range 1 - 5$\cdot10^{-4}$ if in the models the floor is due to collision-induced 
absorption and Rayleigh scattering. If the floor is due to a cloud deck or
haze opacity, then this abundance could be over/underestimated.
Our data alone are not accurate enough to establish the additional contributions of molecules such as CH$_{4}$, CO$_{2}$ and CO detected by other teams 
on the dayside of the planet, or in the terminator. The detection of water vapor is however a sound conclusion.
To date, water vapor absorption in those bands has been reported on two other hot-Jupiters, XO-1b and XO-2b (Tinetti et al. 2010b, Crouzet et al. 2012, Deming et al. 2013) 
and one super-Earth (Berta et al. 2012). 
Our observations do not support the presence of hazes in the near infrared, as highlighted by Figure \ref{water} and the $\chi^2$ test, but they could still be present at shorter wavelengths.

\begin{figure}[f]
 \begin{center}
 \epsscale{1}
 \plotone{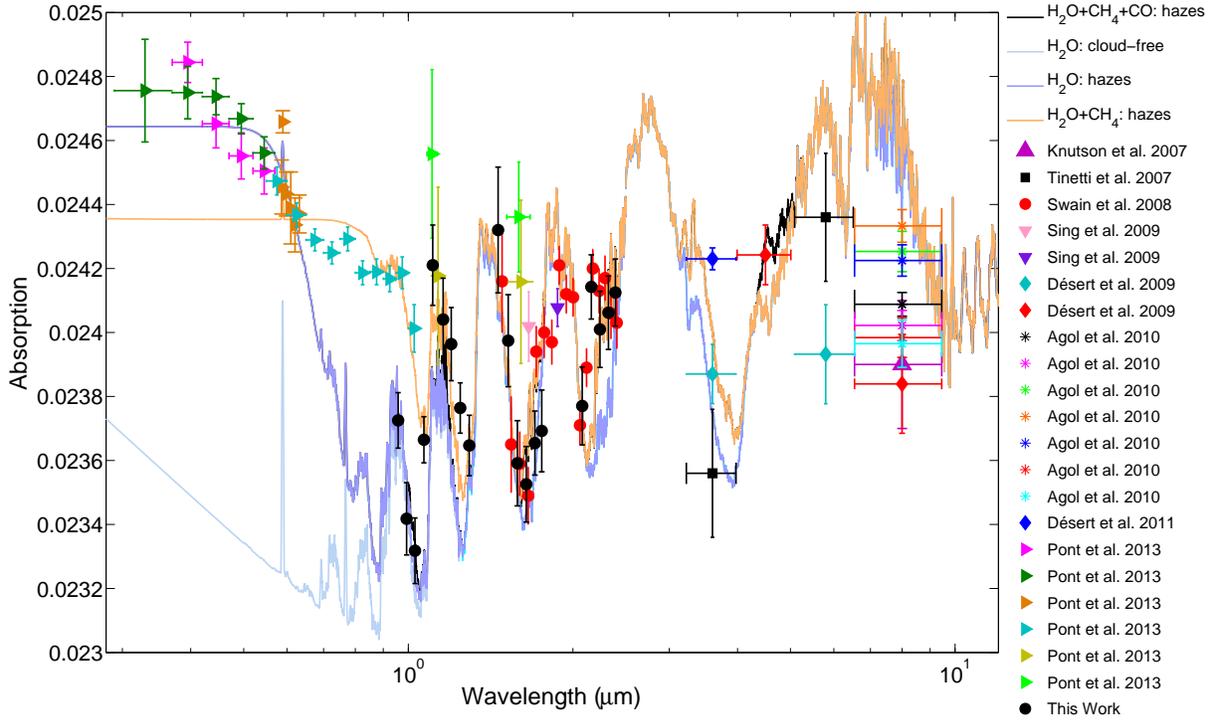}
\end{center}
\caption{0.3 - 24$\mu m$ transmission spectrum of HD-189733b including all the high-precision measurements available in the literature 
(Hubble/STIS - ACS - WFC3 - NICMOS, Spitzer/IRAC from space, IRTF/SpeX from the ground). The data points observed simultaneously are plotted with the same colour. 
We stress that combining multi-epoch datasets is a risky operation: instrumental systematics and
stellar activity may prevent altogether an accurate measurement of the absolute transit depth. Black plot: simulated atmospheric spectrum with water vapor,
methane, carbon dioxide and hazes/clouds. 
Orange plot: modelled spectrum with water vapor, methane and  different haze/cloud contributions. Violet  plot: simulated atmospheric spectrum including only water vapor and hazes/clouds. Light blue plot:  cloud-free spectrum with water vapor.
Note that we plot the newest reanalysis of the STIS, ACS and WFC3 datasets made by the same authors (Pont et al. 2013). 
For the original analyses please refer to Pont et al. 2008, Sing et al. 2011, Huitson et al. 2012, Gibson et al. 2012b. 
Fig. \ref{Nicmos_zoom} shows a zoom-in on the 1.4 - 2.5 $\mu$m range} 
\label{Litplot}
\end{figure}

We plotted our results together with all the other dataset observed with different
instruments and published in the literature (Knutson et al. 2007, Tinetti et al. 2007, Swain et al. 2008, Sing et al. 2009, D\'{e}sert et al. 2009, 2011, 
Agol et al. 2010, Gibson et al. 2012a, Waldmann et al. 2013, Pont et al. 2013).
For this planet transit photometric and spectroscopic data are available from 0.3-24 $\mu$m.
As mentioned in Sect. \ref{previousres}  HD-189733 is a very active star, so there is no guarantee that data recorded at different epochs can be used
to constrain the atmospheric composition (Ballerini et al. 2012).
Furthermore, no currently available instrument can provide an absolute calibration at the level of 10$^{-4}$ of the flux of the star.
With all these caveats we report in Fig. \ref{Litplot} a plausible interpretation of all the available datasets.
We notice that all the IR data are consistent with each other and suggest the presence of molecular features.
We found that contributions from H$_{2}$O, CH$_{4}$ and CO can reasonably match most of the data points.
Data in the UV - VIS look flatter, and show a higher planetary radius compared to the IR; this situation could be explained by the presence of hazes or clouds.

Additional work remains to be done to determine the precise molecular abundances of water and other species.  
To this end, observations with broad, instantaneous spectral coverage will be especially valuable, 
perhaps critical, in order to avoid any significant problem associated with stellar variability.

\section{Conclusions}
We have presented here the first ground-based spectroscopic observations of the primary transit of the hot-Jupiter HD-189733b, recorded with the NASA IRTF/SpeX instrument.
We have pre-cleaned our data and applied the \textit{Model Correlation Fit} technique, finding that the J-, H- and K-band spectra  
are consistent with the collection of datasets recorded from space with Hubble.
The auto-correlation test demonstrates that correlations are efficiently removed, that our residuals are normally distributed and that auto-correlative noise is diminished
at most frequencies.
By comparing the J-, H- and K-12 June 2008 spectra to synthetic models, we found that water vapor with a mixing ratio of around 10$^{-4}$ and 
5$\cdot 10^{-4}$ explains the 
spectral modulations from 0.94 - 2.42 $\mu$m. Our results alone are not sensitive enough to give further constraints on other molecules such as methane, carbon
dioxide or carbon monoxide, as detected in other datasets; however by combining the information contained in other datasets with our results, we can explain the available observations 
with a modelled atmospheric spectrum containing water vapor, methane, carbon monoxide and hazes/clouds.
Future work will involve obtaining broad-band observations taken simultaneously, in order to avoid systematic 
effects resulting from stellar variability and allowing the precise determination of molecular abundances.

The work presented in this paper shows that low-resolution exoplanet spectroscopy is indeed feasible with medium-sized telescopes from the ground. 
While the telluric absorption is a non-negligible hurdle to the sounding of exoplanetary atmospheres, the potential to repeat the observations with relative ease makes
the ground an appealing and complementary option to space.

\newpage

\begin{figure}[f]
 \begin{center}
 \epsscale{1}
 \plotone{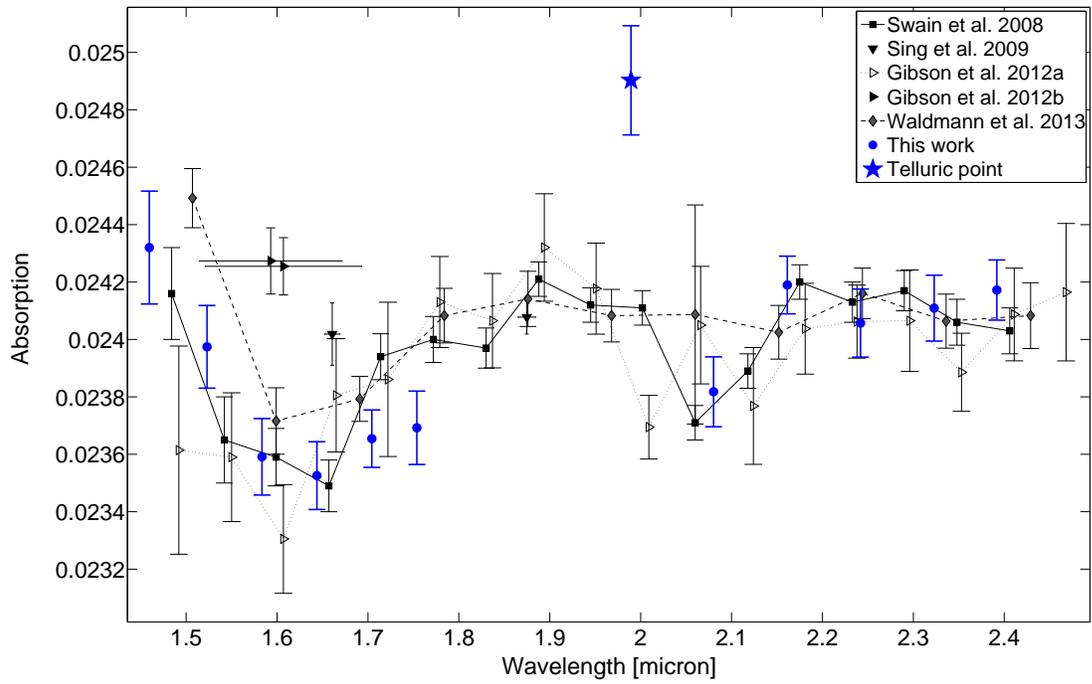}
 \end{center}
\caption{ Zoom-in of all the measurements available for H- and K-bands.}
\label{Nicmos_zoom}
\end{figure}

\begin{table}[tc]

 \centering
 \begin{tabular}{ c c c | c c c }
 \hline
 \hline
 & 2008 June 12 & & & 2008 July 22 & \\
 \hline
 \hline
 $\lambda$ ( $\mu$m) & $\delta$ ($\times10^{-4}$)  & $\varDelta \delta$ ($\times10^{-4}$) & $\lambda$ ( $\mu$m) &$\delta$ ($\times10^{-4}$)  & $\varDelta \delta$ ($\times10^{-4}$) \\ 
 \hline 
  0.958 & 2.733     & 0.866 & -- & --  & --  \\  
  0.993 & -0.343    & 1.130 & -- & --  & --\\
  1.029 &-1.342     & 1.024 & -- & --  & --\\ 
  1.068 & 2.128     & 0.721 & -- & --  & --\\ 
  1.109 & 7.582     & 1.255 & -- & --  & --\\ 
  1.158 & 5.884     & 1.293 & -- & --  & --\\ 
  1.198 & 5.118     & 1.143 & -- & --  & --\\ 
  1.244 & 3.121     & 0.787 & -- & --  & --\\ 
  1.293 & 1.949     & 0.945 & -- & --  & --\\
 \hline 
  1.459 & 5.152 &  1.966 & 1.459 & 4.643 & 1.837\\
  1.523 & 1.694 &  1.439 & 1.538 & -1.602 & 1.213 \\
  1.583 & -2.137 & 1.331  & 1.601 & -0.897& 1.415\\
  1.644 & -2.793 & 1.181  & 1.661 & -0.433 & 1.388\\
  1.704 & -1.506 & 1.000 & 1.721 & -1.843 & 1.549\\
  1.754 & -1.127 & 1.278 &  & & \\
 \hline 
  
  2.080 & -3.035  & 1.218 & 2.099 & -2.180&1.214 \\
  2.161 & 0.685 &   1.004 & 2.180 & 0.350 & 1.248\\
  2.242 & -0.641  &  1.188 & 2.261 & 2.413 & 1.554\\
  2.323 & -0.120 &  1.147 & & & \\
  2.392 & -0.510 & 1.048  & & & \\
 \hline
 \hline
 \end{tabular}
 \caption{Differential depth values and relative errors for the nights of 12 June 2008 (left) and for 22 July 2008 (right).
All the wavelengths are in microns. The telluric points are not listed.}
 \label{depth}
\end{table}





\acknowledgments






\appendix
\section{Noise filtering}
\label{appendix}

\noindent From equation 6 in the manuscript (using a simplified notation here) we have
\begin{equation}
g(t) =  \mathscr{F}^{-1} \left ( \prod_{\lambda=1}^{X} \mathscr{F}[f_{\lambda}(t)] \right ) ^{1/X}
\end{equation}

\noindent Let us add Gaussian white noise to each individual time series $f_{\lambda} (t)$ to get the noisy time series $f_{n,\lambda} (t) = f_{\lambda}(t) + \epsilon_{\lambda}$. We hence have

\begin{align}
g &=  \mathscr{F}^{-1} \left ( \prod_{\lambda=1}^{X} \mathscr{F}[f_{n,\lambda}] \right ) ^{1/X} =  \mathscr{F}^{-1} \left ( \prod_{\lambda=1}^{X} \mathscr{F}[f_{\lambda} + \epsilon_{\lambda}] \right ) ^{1/X}  \\\nonumber
&= \mathscr{F}^{-1} \left ( \prod_{\lambda=1}^{X} \mathscr{F}[f_{\lambda}] + \prod_{\lambda=1}^{X}\mathscr{F} [\epsilon_{\lambda}] \right ) ^{1/X}
\end{align}

\noindent where $(t)$ is omitted for clarity. Since $\epsilon_{\lambda}$ is independent, uncorrelated, stationary Gaussian white noise, we find the geometric mean $\left (\prod_{\lambda=1}^{X}\mathscr{F} [\epsilon_{\lambda}] \right ) ^{1/X}$ to be an exponentially decaying function of $\lambda$.  The distribution as function of $\lambda$ is closely related (but not identical) to a log-normal pdf with $\sigma > 1$ and $\mu = 0$. Such rapid suppression of white noise constitutes the strength of this technique. 
This property is not only limited to white noise (with a constant power spectral density (psd)) but similarly applies to red noise (with 1/f psd) given that the noise realisations are independent.

Through this process, the remaining signal in $g(t)$ constitutes the correlated, non-independent components of $f_{n,\lambda} (t)$. In other words, the science signals and systematics (high and low frequency alike) common to all time series in $X$ are preserved whilst \textquoteleft random\textquoteright ~noise is suppressed nearly regardless of the initial amplitude. The process converges to the common signal floor. \\

This behaviour has two consequences: 

\begin{enumerate}
\item there is no \textquoteleft random\textquoteright ~noise left in $g(t)$ as the white noise components in $f_{n}$ were suppressed. This said, the technique doesn't influence the power-spectral-density distribution of the remaining scatter, which may for the matter behave like white noise and can subsequently be treated as such. Clearly the noise we observe in our data does also show significant low frequency correlations in addition to the high frequency scatter.

\item Once converged, there is little improvement in terms of error-bars by increasing $X$ as the common signal will not diminish and the independent noise component has already converged to near-zero rms. This is in contrast to the more familiar central limit theorem when taking arithmetic means.
 \end{enumerate}





\clearpage

\clearpage


\clearpage



\clearpage




\end{document}